\documentclass[twocolumn,runningheads]{astronepal}
\smartqed
\usepackage{graphicx}
\usepackage{amsmath}
\usepackage{amsfonts}
\usepackage{amssymb}
\usepackage{natbib}

\newcommand{\mps}{\mbox{m\,${\rm s}^{-1}\,$}}

\newcommand{\dps}{\mbox{$^\circ$\,${\rm s}^{-1}\,$}}
\newcommand{\mpa}{\mbox{mag\,${\rm airmass}^{-1}\,$}}

\newcommand\micron{\mbox{$\mu$m}}%

\newcommand\arcdeg{\mbox{$^\circ$}}%
\newcommand\arcmin{\mbox{$^\prime$}}%
\newcommand\arcsec{\mbox{$^{\prime\prime}$}}%

\newcommand\farcmin{\mbox{$.\mkern-4mu^\prime$}}%
\newcommand\farcsec{\mbox{$.\!\!^{\prime\prime}$}}%

\def\asic{Astron. Soc. India Conference Series}
\def\aaps{A\&AS}
\def\basi{Bull. Astron. Soc. India}
\def\cs{Current Science}
\def\fl{Frontline}
\def\icon{IConTOP}
\def\spie{Proc. SPIE}

\begin{document}

\title{Optical astronomical facilities at Nainital, India } 

\titlerunning{Optical astronomy at ARIES, Nainital}  

\author{Ram Sagar, Brijesh Kumar, Amitesh Omar}

\institute{Ram Sagar, Brijesh Kumar, Amitesh Omar\at
              Aryabhatta Research Institute of Observational Sciences\\
             \email{sagar, brij, aomar@aries.res.in}}

\maketitle

\begin{abstract}

~Aryabhatta Research Institute of Observational Sciences (acronym ARIES) 
operates a 1-m aperture optical telescope at Manora Peak, Nainital
since 1972. Considering the need and potential of establishing moderate size optical telescope
with spectroscopic capability at the geographical longitude of India, the ARIES 
plans to establish a 3.6m new technology optical telescope at a new site called 
Devasthal. This telescope will have instruments providing high resolution spectral 
and seeing-limited imaging capabilities at visible and near-infrared bands.  A few other
observing facilities with very specific goals are also being established. A 1.3m
aperture optical telescope to monitor optically variable sources was installed at 
Devasthal in the year 2010 and a 0.5-m wide field (25 square degrees) Baker-Nunn Schmidt 
telescope to produce a digital map of the Northern sky at optical bands was 
installed at Manora Peak in 2011. A 4-m liquid mirror telescope for deep sky survey 
of transient sources is planned at Devasthal. These optical facilities with specialized 
back-end instruments are expected to become operational within the next few years and 
can be used to optical studies of a wide variety of astronomical topics including 
follow-up studies of sources identified in the radio region by GMRT and UV/X-ray 
by ASTROSAT. 


\end{abstract}


\section{Introduction}

\begin{figure*}
\centering
\includegraphics[width=17cm]{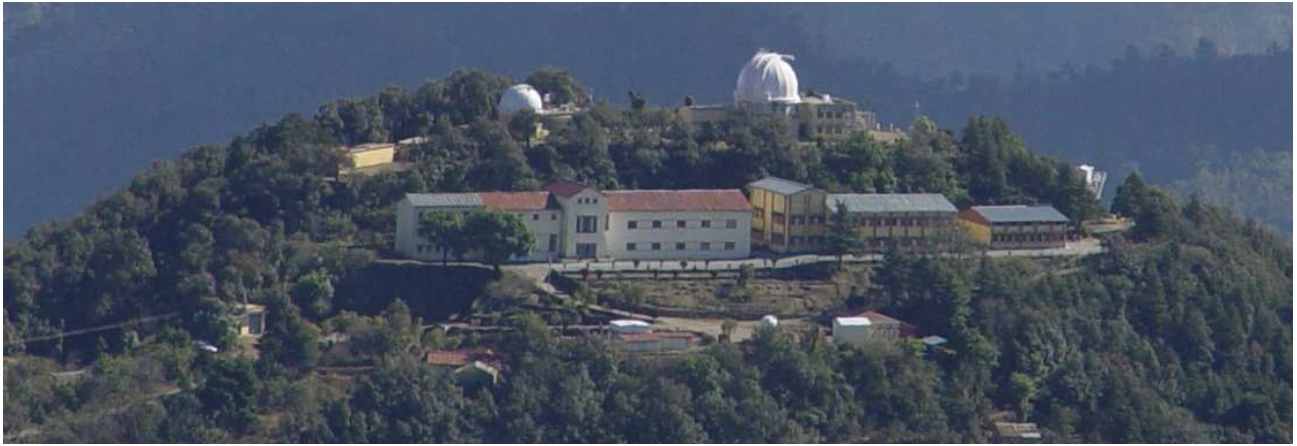}%
\caption{A panoramic view of Aryabhatta Research Institute of Observational 
         Sciences (acronym ARIES) located at Manora Peak, Nainital. The larger white dome houses
         104-cm Sampurnanand Optical Telescope which is in operation since 1972.}
\label{fig:manora}
\end{figure*}

Aryabhatta Research Institute of Observational Sciences (acronym ARIES), an
autonomous research institute under the Department of Science and Technology, 
Government of India, is located on Manora Peak near the city of 
Nainital (Figure~\ref{fig:manora}). The institute builds and operates observational 
facilities to carry out frontline research in the areas of Astrophysics and Atmospheric 
Physics. The institute came into existence on 20th April 1954 as an astronomical 
observatory under the state government of Utter Pradesh, India. 
On 22nd March 2004, the administrative control of the State Observatory was taken over 
by the Department of Science and Technology, Government of India and it was renamed 
as ARIES, to signify the location of Sun in the zodiac 
ARIES \citep{2004FL.21..30P,2006BASI...34...37S} at the epoch of its formation in 1954 
and reincarnation in 2004. The present contribution focus
on optical astronomy and it gives an overview of the existing observational facilities 
at ARIES as well as the new initiatives taken up during last decade. The need and importance 
of new observing facilities in optical astronomy is also described.  


\section{Manora Peak} \label{sec:manora}  

The Manora Peak\footnote[1]{Longitude : 79\arcdeg27\arcmin25\farcsec5 E, 
Latitude : 29\arcdeg21\arcmin39\farcsec0 N, Altitude : 1927 m} gets about 280 cm of rain annually,
of which about 230 cm is concentrated during monsoon months from July to September. The
meteorological data indicate that about 150 photometric and 200 spectroscopic nights are 
expected annually. The seeing at Manora Peak is usually better than 2\arcsec. 
The study of extinction properties over 35 years 
since 1970 \citep{2000BASI...28..675S} indicates that the photometric quality of nights at 
Manora Peak are stable and there is no noticeable aerosol contamination of the sky. The mean 
atmospheric extinction values at Manora Peak are 0.57, 0.28, and 0.17
\mpa in the U, B and V bands respectively, while the corresponding best observered values 
are 0.45, 0.20 and 0.10 \mpa. The observational facilities at Manora Peak are described 
below.

\subsection{The 0.5m Schmidt telescope}

During the International Geophysical year (1957-58), a 79/51-cm f/1 Baker-Nunn
satellite tracking camera was installed at the Institute by the Smithsonian
Astrophysical Observatory, USA. It was the only center in India but actively 
networked as a part of the 12 centers established all over the globe. The first
photograph of an artificial satellite was taken on 29th August 1958. The camera
successfully photographed a total of over 45,700 satellite transits. After
1976, the camera is not in use due to the advent of modern observational
techniques in the area. Following successful conversion of such cameras
into a wide field Schmidt-telescopes for carrying out astronomical survey work
by Australian and Spanish group, ARIES initiated this job in 2005.

The basic optical design of the 79/51-cm f/1 Baker-Nunn satellite tracking
camera uses a three-element corrector to produce an extremely wide field of
view across a curved plane at the prime focus for photographic imaging.
Major jobs in converting the existing Baker-Nunn camera into a 0.5-m Schmidt
telescope with CCD imaging capabilities are (i) modification of optical design
from photographic curved to flat focal plane for CCD observations, (ii)
changing the mounting system from alt-azimuthal to Equatorial English mount
(iii) computer control of the telescope, and (iv) optical alignment and
installation of a new customized CCD imaging system at the prime focus
having plate scale of $\sim 7$ arcmin per mm. Further technical details 
can be found elsewhere\citep{2009icon...1..381S}

\begin{figure}
\centering
\includegraphics[width=8.4cm]{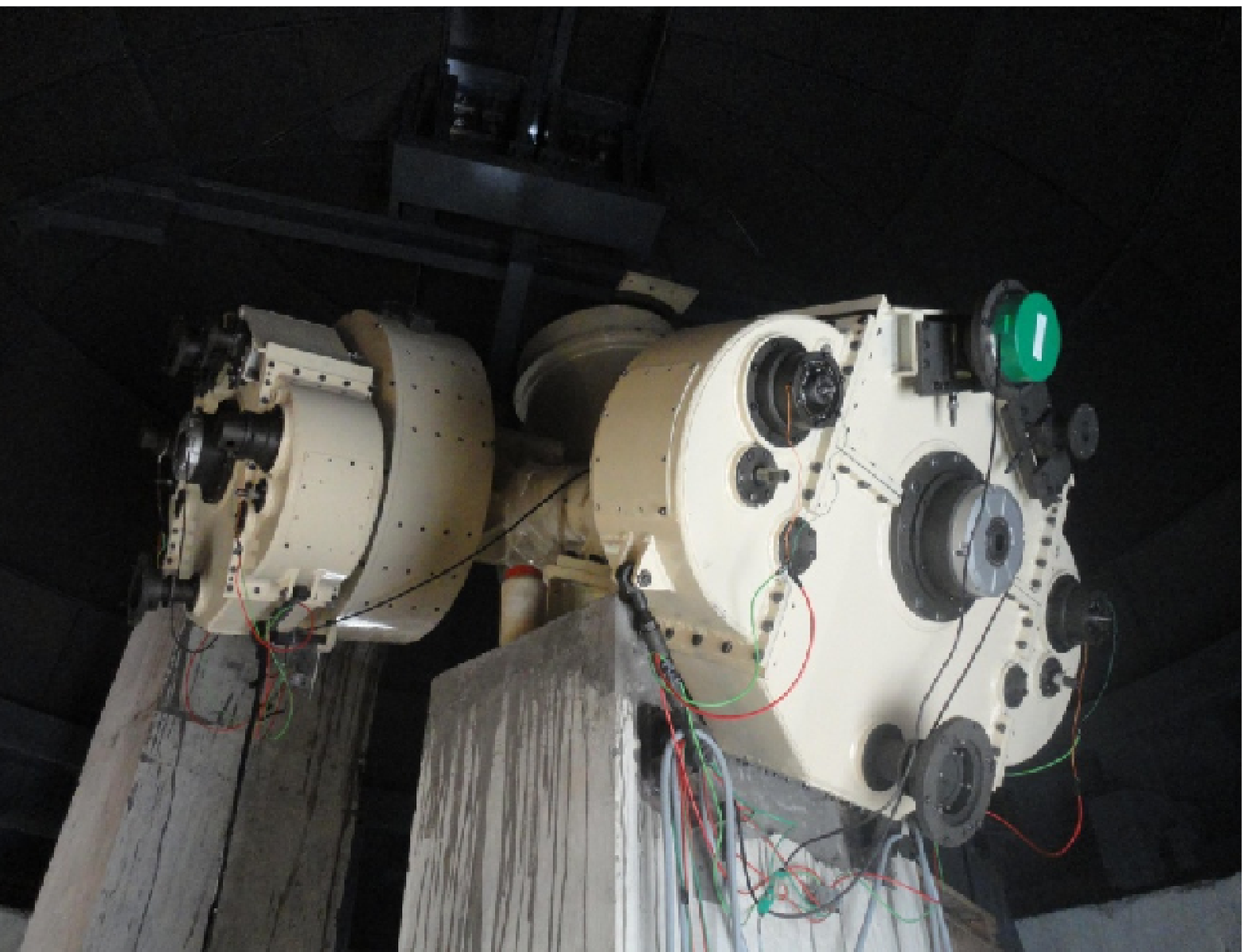}\\%
\includegraphics[width=8.4cm]{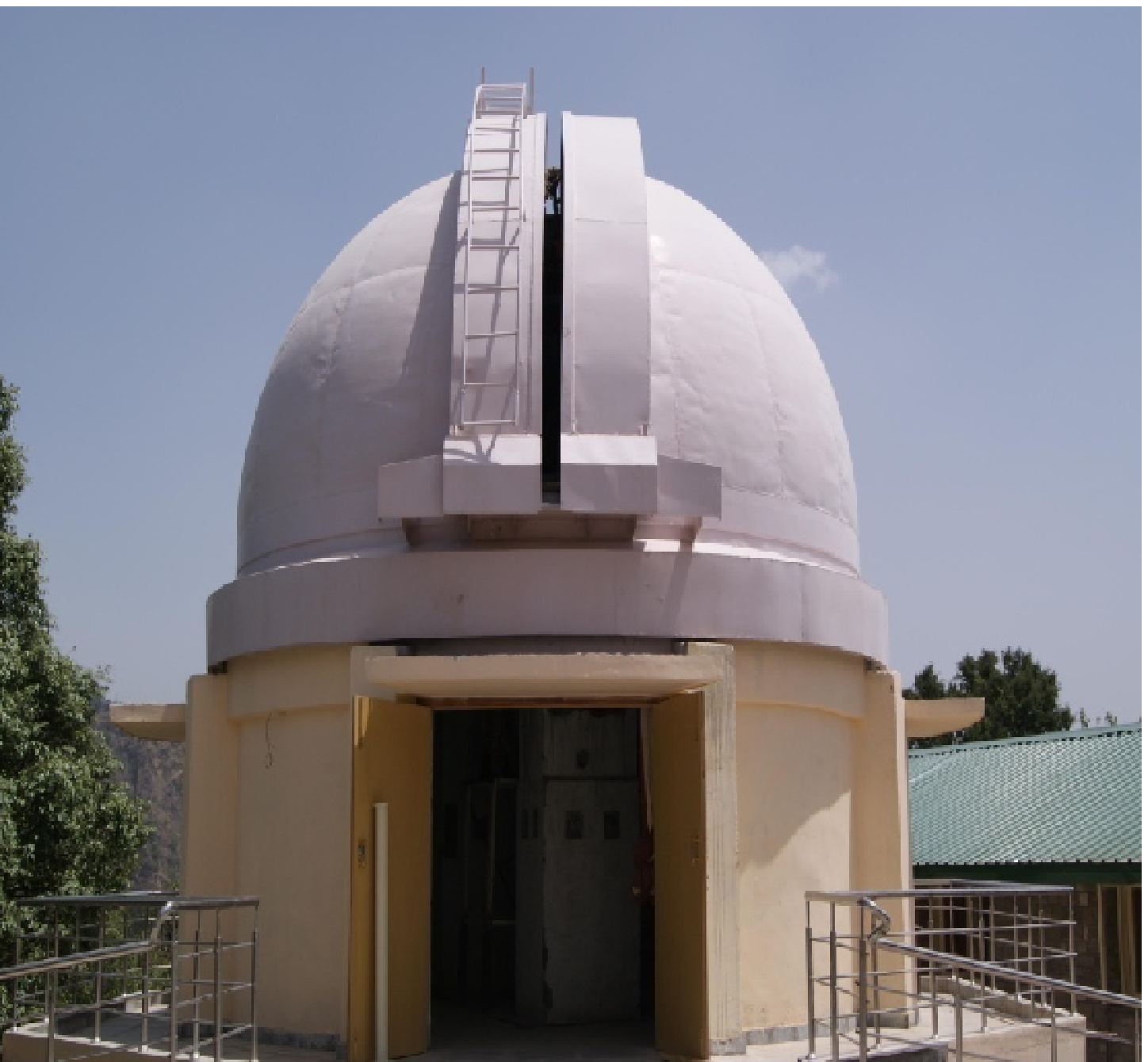}%
\caption{The 0.51-m Schmidt telescope at Manora Peak. The telescope is shown on top and the 
         dome is shown in bottom panel.}
\label{fig:bnc}
\end{figure}

The telescope has been successfully installed at Manora Peak in 2011
(see Figure~\ref{fig:bnc}) and currently the alignment and fine tuning
is in progress. The dome control system has been designed and developed
in-house at ARIES. The dome can automatically synchronize with the telescope
motion. The computer simulation suggests that the Schmidt camera 
can reach 20th magnitude with 10\% photometric accuracy for an integration
time of 1 min. Scientific programs like study of variable stars, Asteroids and
Near-Earth-Objects, detection of extra-solar planets though transit method,
transient objects like GRBs and supernovae, and imaging of large star
clusters could be suitably accomplished with this wide-field imaging
telescope.


\subsection{The 104-cm Sampurnanand Telescope}
\begin{figure}
\centering
\includegraphics[width=8.4cm]{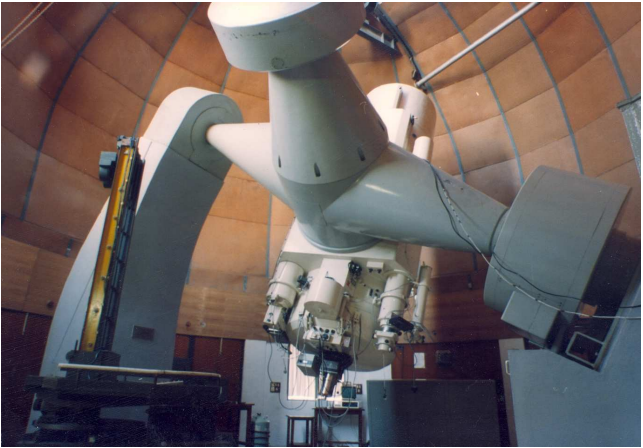}%
\caption{The 104-cm Sampurnanad Optical Telescope at Manora Peak, Nainital.}
\label{fig:st}
\end{figure}

The 104-cm Sampurnanand Telescope was supplied by Veb Carl Zeiss, Jena and installed at Manora 
Peak in 1972. The telescope is
a Ritchey-Chretien reflector with a f/13 Cassegrain and f/31 Coude focii. It has an 
equatorial 2-pier English mount (see Figure~\ref{fig:st}). A field of 45 arcmin diameter 
is available with corrector at the Cassegrain focus. During seventies and eighties, the telescope
was primarily equipped with a photoelectric photometer and a spectrum scanner. Further
technical details can be found in \citet{2006BASI...34...65S}. The guiding of the telescope 
is done using an ST4 camera mounted with the 8-inch guider telescope and a long
exposure (30-min) can be given without trailing. A detailed description of the 
telescope can be found in \citet{1999CS..77..643P}. Currently, the telescope is equipped
with a 2k$\times$2k CCD camera which covers around 13 arcmin of sky, an imaging polarimeter
and a three channel fast photometer. Over 650 scientific publications and 40 PhD Thesis
have been produced utilizing observational data taken from this telescope.    


\section{Multi-wavelength observations and role of optical astronomy} \label{sec:opt}
The observations of celestial objects at multi-wavelengths are essential to establish their 
identity and their meaning in astrophysical terms as they radiate across the entire 
electromagnetic spectrum. For such observations, India has undertaken major initiatives recently, 
and plans to establish front line observational facilities in future too. For example, 
Giant Meter-wave Radio Telescope (GMRT) installed during the last decade and the upcoming 
India's first Multi-wavelength Astronomical Satellite, ASTROSAT are world-class Indian 
Observing Facilities at radio, X-ray and ultra-violet wavelengths. However, the observing 
facilities at optical wavelength are far from being world class as all the existing optical 
telescopes in India\footnote[2]{The existing optical telescope in India includes 2-m Himalayan
Chandra Telescope, Hanle; 1.2-m telescope, Mt. Abu; 1.2-m telescope Japal Rangapur, Hyderabad; 
2.34-m VBT, Kavalur; 2-m telescope IGO, Girawali near Pune.} are less than 2.34-m aperture in size while 
in the world moderate 
(3 to 6 m) size telescopes were established about 3 to 4 decades ago and large (8 to 16 m) 
ones during the last decade. Now-a-days a number of large ($>$ 8 m) optical telescopes are 
under fabrication in different parts of the globe. It becomes therefore vital for India 
to set up moderate size optical telescopes during coming years and plan for larger 
size optical telescopes in future. Technological advancements and the availability of sensitive 
detectors have made moderate size optical telescopes not only economical but also extremely 
valuable even today due to the increased level of performance and minimal maintenance for 
such systems \citep{2000CS..78.1076P}.


\begin{figure}
\centering
\includegraphics[width=8.4cm]{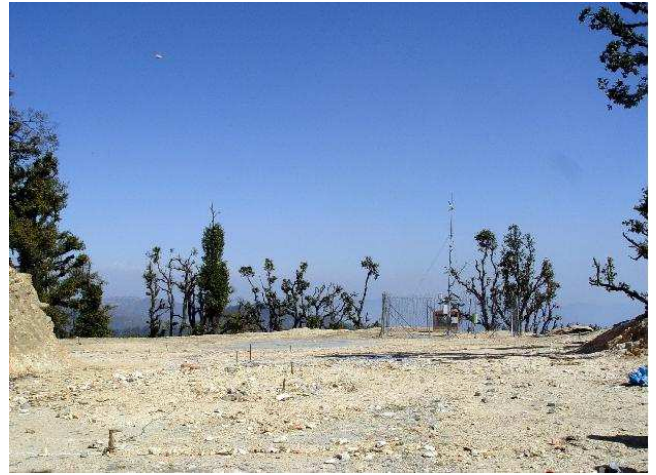}%
\caption{The proposed site at Devasthal for the 3.6m optical telescope.}
\label{fig:dev}
\end{figure}


\section{Devasthal} \label{sec:opt}

An extensive site characterization conducted during 1980 - 2001 in the central
part of the Himalayan Range, identified Devasthal
(Lat:29\arcdeg 21\arcmin40\arcsec N, Lon:79\arcdeg41\arcmin04\arcsec E,
Alt: 2450 m above msl) as a potential site for
optical observations \citep{1999A&AS..136...19P, 2000A&AS..144..349S, 2001BASI...29...39S}. 
Seeing measurements close to ground level were successfully
carried out on 80 nights during 1998-99 using modern differential image motion monitor.
The results indicate a median ground level seeing
estimate of about 1\farcsec1; the 10 percentile values between 0\farcsec7\,
to 0\farcsec8\, (mean = 0\farcsec75) while for 35\% of the time the
seeing was better than 1\arcsec. Microthermal measurements indicate that if the telescope
can be installed above 8 m above from the ground, the seeing could be sub-arcsec. 
These coupled with the number of yearly
spectroscopic nights ($\sim$ 210), darkness of the per square arcsec sky
(V $\sim$ 21.8 mag) and other atmospheric parameters for Devasthal make this
site comparable to the international standards\citep{2000A&AS..144..349S, 2001BASI...29...39S} .
Devasthal site (see Figure~\ref{fig:dev}) is located about 50 km by road east of Nainital having a direct
line-of sight distance ($\sim$ 22 km) from the present location of ARIES at
Manora Peak. The upcoming optical facilities at Devasthal are described below.

\subsection{The 1.3m optical telescope} \label{sec:dfot}

\begin{figure*}
\centering
\includegraphics[width=6cm]{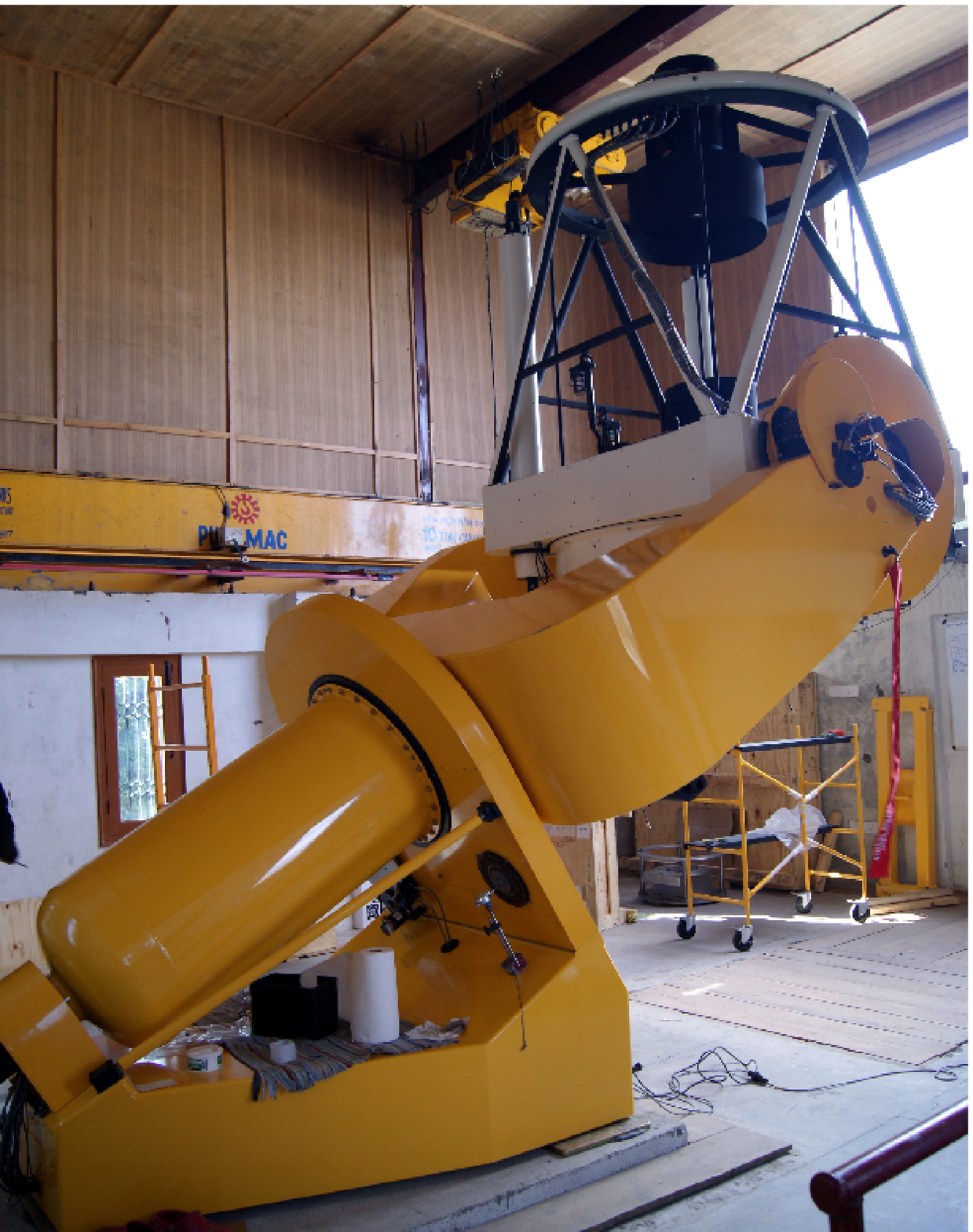}%
\includegraphics[width=5.3cm]{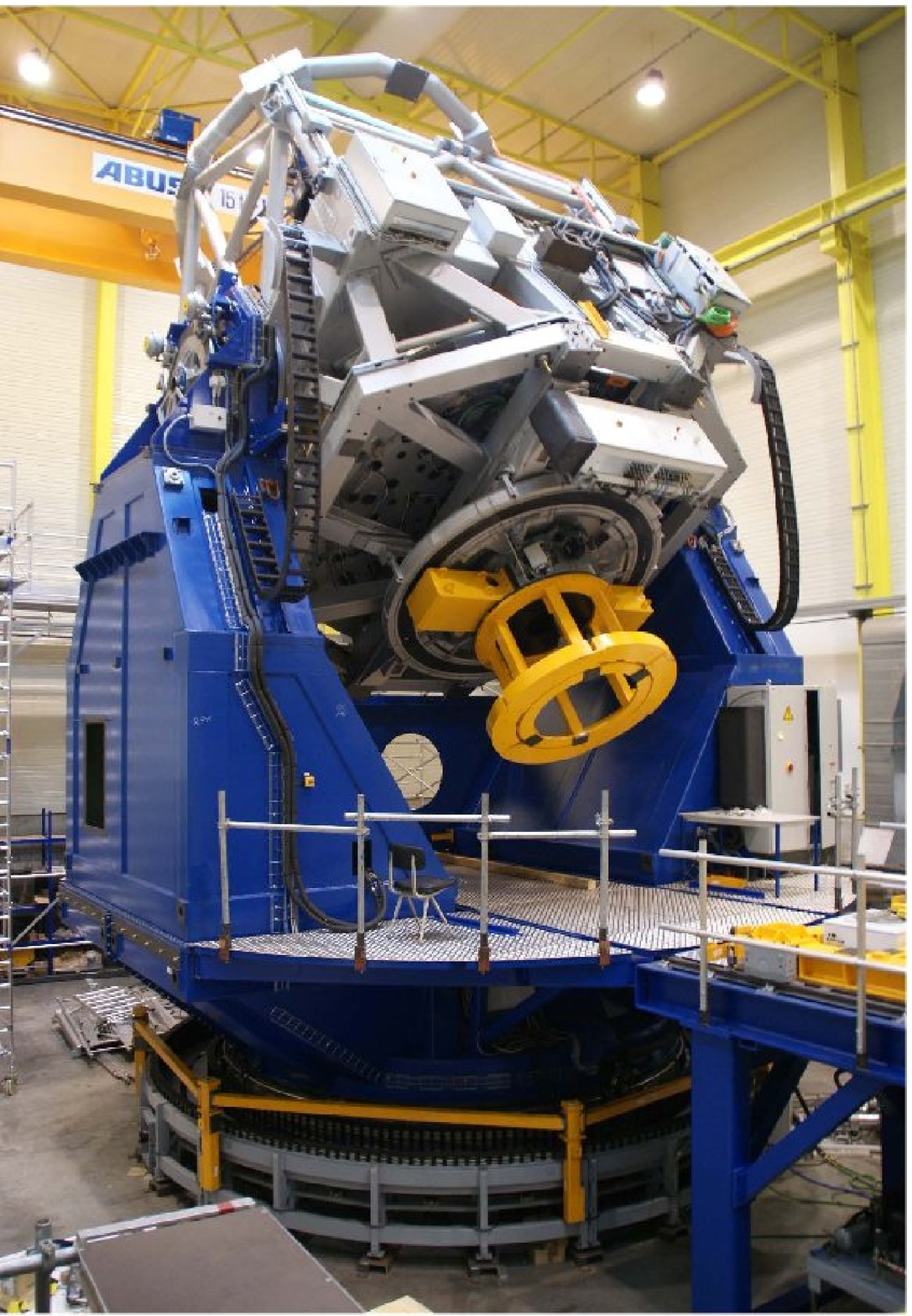}%
\includegraphics[width=6cm]{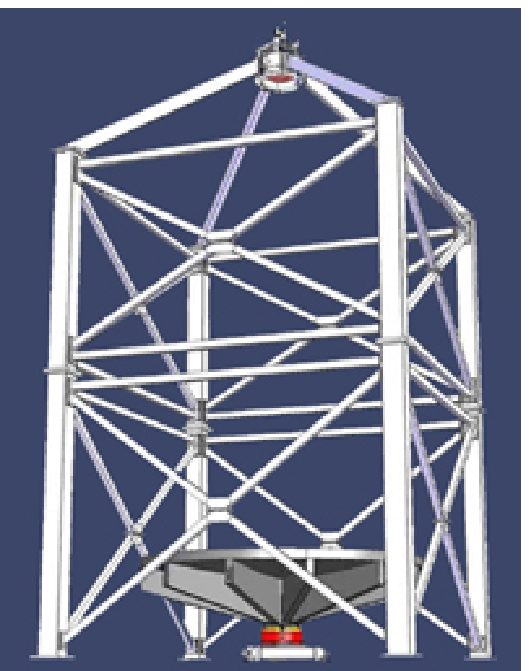}%
\caption{The 1.3m, f/4 Optical Telescope, installed at Devasthal site (leftmost), the 3.6m, f/9 
         optical telescope, as-built system at the AMOS workshop, Belgium (middle), and 
         the 4-m, f/2 International Liquid Mirror Telescope, computer rendering of the 
         design (rightmost).}
\label{fig:tel}
\end{figure*}

In October 2010, a new modern 1.3-m Devasthal Fast Optical Telescope (DFOT) has been
installed successfully at Devasthal \citep{2011CS.101.1020P}. A picture of the
telescope is shown in Figure~\ref{fig:tel}. In order to avoid degradation 
of seeing due to local environments, the telescope is mounted 3 m above the ground and the 
enclosure has a roll-off roof design. The telescope design of the 2-mirror
Ritchey-Chr{\'e}tien optics along with a single element corrector is optimized to deliver 
a fast beam (f/4, plate scale of 40\arcsec\,mm$^{-1}$) and a naturally flat-field of 
66\arcmin\,diameter at the axial Cassegrain focus \citep{2000SPIE.4003..456M}\,. It is therefore
suitable for wide-area survey of a large number of point as well as extended sources.
Without autoguider the tracking accuracy of the telescope is better than 0\farcsec5
in an exposure of 300 s up to a zenith distance of 40\arcdeg. The pointing accuracy
of the telescope is better than 10\arcsec\,Root Mean Square (RMS) for any point in the sky. 
Further technical details on the as-designed specifications of the telescope system are given
elsewhere \citep{2010ASInC...1..203S,2012ASInC...4..113S}\,. The main scientific
objective is to monitor optical and near infrared (350-2500 nm) flux variability
in the astronomical sources such as transient events (Gamma-ray bursts, supernovae),
episodic events (active galactic nuclei, X-ray binaries and cataclysmic variables), 
stellar variables (pulsating, eclipsing and irregular), transiting extrasolar planets - and to 
carry out photometric and imaging surveys of extended astronomical sources, e.g. HII regions,
star clusters, and galaxies. Further details on the scientific objectives can be
found elsewhere \citep{2006BASI...34...37S}.

A differential light curve of the WASP-12 transiting system along with the model fit indicates 
a photometric precision of 1 mmag for a 11.7 mag star. As a comparison, a similar observations 
using 104-cm Sampurnanand Telescope at Manora Peak, we get an accuracy of about 3 to 4 mmag. 
Hence the 1.3-m DFOT at Devasthal would be suitable for the scintillation limited science 
programs requiring a detection of few mmag on a time scale of hrs (e.g. exoplanet search 
and AGN variability).

A detailed report on commissioning of the 1.3-m DFOT can be found 
elsewhere\citep{2012ASInC...4..113S}\,.


\subsection{The 3.6-m Optical Telescope}

The 3.6-m optical telescope is a Ritchey-Chretien f/9 system. The primary mirror 
is made from Schott zerodur glass
and it is a concave hyperboloid f/2 while the secondary is made from 
Astrositall glass and it is a convex hyperboloid f/2.6. The telescope 
has Cassegrain focus fitted with a 30\arcmin\, wide field three-lens corrector,
auto-guiding unit and a derotator instrument interface. The telescope has
two side and one main ports (see Figure~\ref{fig:tel}). 

The telescope has an alt-azimuth mount and the bearings of the telescope uses
latest technology available in the market. The tracking performance of the telescope is better
than 0\farcsec1 RMS for one minute in open loop with winds inside the dome of less
than 3 \mps\, and in close loop it is 0\farcsec11 for less than one hour. In open loop and for wind
inside the dome of 5 \mps\, the tracking accuracy is $\sim$0\farcsec5 peak in 15 minute.
The maximum selwing speed of the telescope is 2\dps\, and 1\dps\, in azimuth 
and altitude respectively.

The optics of the telescope are designed and polished to deliver images with 
encircled energy of 80\% in less than 0\farcsec45 diameter in a 10\arcmin\, arcmin 
Field of View (FoV) over 350 nm to 1500 nm wavelength range without corrector.   
A preliminary speckle imaging tests performed at the AMOS workshop suggest that 
the as-built optics can deliver images with E80 better than 0\farcsec3
diameter \citep{2012SPIE.8444..102F}.

The telescope will be housed in a cylindrical dome like structure having 
height of 21.5 m  and a diameter of 16.5 m. In order to avoid degradation of 
seeing due to dome, two separate ventilation ducts, one from telescope floor
and another from telescope technical room have been provided. An auxiliary 
building housing aluminising plant from mirrors up to 3.7m diameter have
alos been envisioned. More details on the telescope dome and the auxiliary 
building can be found elsewhere \citep{2012SPIE.8444..152F}. 

The first generation focal plane instruments are a Faint Object Spectrograph
and Camera (FOSC) and a CCD optical imager. The second generation instruments
include an optical near-infrared spectrograph and imager; a high resolution 
optical spectrograph and an integral field unit.

The Faint Object Spectrograph and Camera (FOSC) is a focal reducer instrument. The 
instrument will have imaging capabilities with one pixel resolution of less 
than 0\farcsec2 in the FoV of $\sim$ 14\arcmin\, $\times$ 14\arcmin\, of the telescope, and
low-medium spectroscopy with spectral resolution (250-4000)
covering the wavelength range from 350 nm to 900 nm. A computer simulation indicate
that we can image a 25 mag star in $V$ band with an hour of exposure time.
The optical and mechanical design of the instrument has been completed in-house at
ARIES. Further technical details can be found elsewhere \citep{2012SPIE.8446..38F}\,.

An optical imager with a 15\micron\, pixel of square size, 4k$\times$4k back-illuminated 
CCD detector, liquid
nitrogen cooling, full frame window mode operation, and the associated control electronics has
also been proposed as a first light instrument. A contract for assembly and integration of the
CCD camera has been awarded to Semiconductor Technology Associates,
USA\footnote[3]{http://www.sta-inc.net/}. The mechanical interface for the camera is being
designed and manufactured in-house at ARIES. This imager will primarily be used to verify the
performance of telescope during the commissioning phase. The imager will cover a square
area of 6\farcmin5$\times$6\farcmin5 on the sky. This instrument will have broadband
Johnson-Cousins $UBVRI$ and $ugriz$ SDSS filters, as well as a few narrow-band filters. 

The above mentioned focal plane instruments shall be used to carry out
observations for the studies related to exo-planets, stellar variability and
asteroseismology, interacting binary systems, variability in latest type
soft x-ray stars, formation and evolution of stars, studies of galaxies, dark
matter in the galaxy, optical follow-up of the sources identified by
GMRT and ASTROSAT and the highly energetic events - SNe and GRBs.

\subsection{The 4-m International Liquid Mirror Telescope}
The 4-m International Liquid Mirror Telescope (LMT) uses Liquid Mirror Technology and 
the mercury mirror of the telescope 
will have a diameter of 4 m and a focal length of 8 m (see Figure~\ref{fig:tel}).
The ILMT is proposed to be installed at Devasthal as a joint collaboration between 
India, Belgium and Canada. It will perform as a transit telescope. A 4k$\times$4k CCD detector
with a suare size pixel of 15 \micron\, shall be positioned at the prime
focus of the telescope and it will cover an area of about 30\arcmin\,$\times$30\arcmin\, on the sky. 
The mirror being parabolic in shape needs a corrector to get a
flat focal surface of about 30\arcmin\, diameter. The rotation of the Earth induces the motion
of the sky across the detector surface. The CCD detector works in a
time delay integration mode, i.e. it tracks the stars by electronically
stepping the relevant charges at the same rate as the target moves
across the detector, allowing the integration as long as the target
remains inside the detector area. At the latitude of Devasthal,
a band of half a degree covers 156 square degrees out of which 88 square degrees
being covered at high galactic latitude (b $>$ 30\arcdeg) including the
direction of the north galactic pole. The nightly
integration times are rather short, typically 120 s but it is possible
to co-add data from selected nights in order to get sky images of longer
integration times. About 10 Gigabytes of data will be collected each night.

The expected 5$\sigma$ limiting magnitudes achieved by co-adding scans are 24.5 at $U$, $B$ and $V$ 
bands, 23.5 at $R$ and $I$ bands and 22.3 at Gunn-z band. The expected database towards the Galactic Bulge
direction includes 10 million stars, 30000 variables, 8000 binaries,
8000 LPVs/SRVs, 5000 spotted RSCVn, 1400 RR Lyrae, 250 $\delta$-Scuti,
20 Cepheids, 50 yr$^{-1}$ microlenses and 5 yr$^{-1}$ Cataclysmic variables - providing
valuable inputs for the studies of stars, galaxies and cosmology. More details 
on the 4-m ILMT can be found elsewhere\footnote[4]{http://www.aeos.ulg.ac.be/LMT/}.


\section{Summary}
The technological advancements and the availability of sensitive detectors has made moderate size, 
2-m to 4-m class, optical telescopes extremely valuable even today. Furthermore, such telescopes have 
advantages over large ones in availability, survey and follow -up work. In next few years
a 3.6-m optical telescope with active mirror technology and another 4-m telescope with liquid
mirror technology will become operational at Devasthal. All these facilities at Devasthal will
be valuable addition to the existing optical astronomical facilities in India as well as in the globe. 

\bigskip

\textbf{Acknowledgements} 
The work presented here is on behalf of a larger team associated with
the development of Devasthal Observatory. The authors are thankful to the Devasthal
Observatory staff for their assistance during observations taken with 1.3-m telescope. 
Approvals and encouragements to initiate the above projects from all the members 
of the Governing Council of ARIES are gratefully acknowledged. Useful scientific inputs from 
members of ARIES project team are thankfully acknowledged.\\


\begin{thebibliography}{}

\bibitem[Kumar et al.(2000)]{2000BASI...28..675S}
{Kumar}, B., {Sagar}, R., {Rautela}, B.~S., {Srivastava}, J.~B., {Srivastava}, R.~K.,
   ``{Sky Transparency over Nainital : A retrospective study},'' {\em
  \basi}~{\bf 28},  675 (2000).

\bibitem[Melsheimer \& MacFarlane(2000)]{2000SPIE.4003..456M}
{Melsheimer}, F.~M. and {MacFarlane}, M.~J., ``{Very wide field, very fast
  telescope},'' in [{\em Optical Design, Materials, Fabrication, and
  Maintenance}{\nolinebreak\hspace{0.1em}]},  {Dierickx}, P., ed., {\em \spie}
  {\bf 4003},  456--463 (2000).


\bibitem[Mondal et al.(2009)]{2009icon...1..381S}
{Mondal}, S., {Gupta}, K.~G., {Lata}, S., {Medhi}, B.~J., {Bangia}, T., {Kumar}, T.~S.,
  {Yadav}, S., and {Singh}, S.~K., ``{Development of ARIES Baker-Nunn camera to a
  wide-field imaging telescope with CCD},'' in [{\em Tends in optics and 
  photonics}{\nolinebreak\hspace{0.1em}]},
  {Ghosh}, A., ed., {\em \icon} {\bf 1},  381--386 (2009).

\bibitem[Ninane et al.(2012)]{2012SPIE.8444..102F}
{Ninane}, N., {Bastin}, C., {de Ville}, J., {Michel}, F.~J., {Pierard}, M.,
  {Gabriel}, G., {Flebus}, C., and {Omar}, A., ``{The 3.6 m Indo-Belgian
  Devasthal Optical Telescope: assembly, integration and tests at AMOS},'' in
  [{\em Ground-based and Airborne Telescopes IV}{\nolinebreak\hspace{0.1em}]},
  {\em \spie} {\bf 8444} (2012).

\bibitem[Omar et al.(2012)]{2012SPIE.8446..38F}
{Omar}, A., {Yadav}, R., {Shukla}, V., {Mondal}, S., and {Pant}, J., ``{Design
  of FOSC for 360-cm Devasthal Optical Telescope},'' in [{\em Ground-based and
  Airborne Instrumentation for Astronomy IV}{\nolinebreak\hspace{0.1em}]},
  {\em \spie} {\bf 8446} (2012).

\bibitem[Pandey et al.(2012)]{2012SPIE.8444..152F}
{Pandey}, A., {Shukla}, V., {Bangia}, T., {Rasker}, R., and {Kulkarni}, R.,
  ``{Enclosure design for the ARIES 3.6m optical telescope},'' in [{\em
  Ground-based and Airborne Telescopes IV}{\nolinebreak\hspace{0.1em}]},  {\em
  \spie} {\bf 8444} (2012).

\bibitem[Pant et al(1999)]{1999A&AS..136...19P}
{Pant}, P., {Stalin}, C.~S., and {Sagar}, R., ``{Microthermal measurements of
  surface layer seeing at Devasthal site},'' {\em \aaps}~{\bf 136},  19--25 (1999).

\bibitem[Ramachandran(2004)]{2004FL.21..30P}
{Ramachandran}, R., ``{An Institute reborn},'' {\em \fl}~{\bf 21},  30--32
  (2004).

\bibitem[Sagar(1999)]{1999CS..77..643P}
{Sagar}, R., ``{Some new initiatives in optical astronomy at UPSO, Nainital},''
  {\em \cs}~{\bf 77},  643--652 (1999).

\bibitem[Sagar et al.(2000)]{2000A&AS..144..349S}
{Sagar}, R., {Stalin}, C.~S., {Pandey}, A.~K., {Uddin}, W., {Mohan}, V.,
  {Sanwal}, B.~B., {Gupta}, S.~K., {Yadav}, R.~K.~S., {Durgapal}, A.~K.,
  {Joshi}, S., {Kumar}, B., {Gupta}, A.~C., {Joshi}, Y.~C., {Srivastava},
  J.~B., {Chaubey}, U.~S., {Singh}, M., {Pant}, P., and {Gupta}, K.~G.,
  ``{Evaluation of Devasthal site for optical astronomical observations},''
  {\em \aaps}~{\bf 144},  349--362 (2000).


\bibitem[Sagar(2000)]{2000CS..78.1076P}
{Sagar}, R., ``{Importance of small and moderate size optical telescopes},''
  {\em \cs}~{\bf 78},  1076--81 (2000).

\bibitem[Sagar(2006)]{2006BASI...34...37S}
{Sagar}, R., ``{Aryabhatta Research Institute of Observational Sciences:
  reincarnation of a 50 year old State Observatory of Nainital},'' {\em
  \basi}~{\bf 34},  37 (2006).

\bibitem[Sagar et al.(2010)]{2010ASInC...1..203S}
{Sagar}, R., {Kumar}, B., {Omar}, A., and {Pandey}, A.~K., ``{New initiatives
  in optical astronomy at ARIES},'' in [{\em Interstellar Matter and Star
  Formation: A Multi-wavelength Perspective}{\nolinebreak\hspace{0.1em}]},
  {Ojha}, D.~K., ed., {\em \asic} {\bf 1},  203--210 (2010).

\bibitem[Sagar et al(2011)]{2011CS.101.1020P}
{Sagar}, R., {Omar}, A., {Kumar}, B., {Maheswar}, G., {Pandey}, S.~B.,
  {Bangia}, T., {Pant}, J., {Shukla}, V., and {Yadava}, S., ``{The new 130-cm
  optical telescope at Devasthal, Nainital},'' {\em \cs}~{\bf 101},  1020--23 (2011).

\bibitem[Sagar et al.(2012)]{2012ASInC...4..113S}
{Sagar}, R., {Kumar}, B., {Omar}, A., and {Joshi}, Y.~C., ``{New optical
  telescopes at Devasthal Observatory : 1.3-m installed and 3.6-m upcoming},''
  in [{\em Recent advances in observational and theoretical studies of star
  formation}{\nolinebreak\hspace{0.1em}]},  {Subramaniam}, A., and {Anathpindika}, 
  eds., {\em \asic} {\bf 4}, 113--120 (2012).

\bibitem[Sinvhal et al.(1972)]{1972oams.conf...20S}
{Sinvhal}, S.~D., {Kandpal}, C.~D., {Mahra}, H.~S., {Joshi}, S.~C., and
  {Srivastava}, J.~B., ``{The 104-cm telescope of Uttar Pradesh State
  Observatory},'' in [{\em Optical astronomy with moderate size telescopes,
  Symposium held at Centre of Advanced Study in Astronomy, Osmania Univ.,
  Hyderabad}{\nolinebreak\hspace{0.1em}]},   20--34 (1972).

\bibitem[Sinvhal(2006)]{2006BASI...34...65S}
{Sagar}, R., ``{The Uttar Pradesh State Observatory - some recollections
  and some history (1954-1982)},'' {\em \basi}~{\bf 34},  65 (2006).

\bibitem[Stalin et al.(2001)]{2001BASI...29...39S}
{Stalin}, C.~S., {Sagar}, R., {Pant}, P., {Mohan}, V., {Kumar}, B., {Joshi},
  Y.~C., {Yadav}, R.~K.~S., {Joshi}, S., {Chandra}, R., {Durgapal}, A.~K., and
  {Uddin}, W., ``{Seeing and microthermal measurements near Devasthal top},''
  {\em \basi}~{\bf 29},  39--52 (2001).




\end{thebibliography}
\end{document}